\documentclass[aps,prb,preprint,amsfonts,showpacs,letterpaper]{revtex4-1}
\usepackage{graphicx}
\usepackage{hyperref}

\def \fo       {Fe$_3$O$_4$}

\begin{document}

\title{First-principles determined charge and orbital interactions in Fe$_3$O$_4$}

\author{Fei Zhou}
\email{fzhou@ucla.edu}
\altaffiliation{Now at University of California, Los Angeles}
\author{Gerbrand Ceder}
\email{gceder@mit.edu}
\affiliation{Department of Materials Science and Engineering, Massachusetts Institute of Technology, Cambridge, MA 02139, USA}

\date{\today}
\pacs{71.10.Li, 71.28.+d, 75.30.Et}

\begin{abstract}
The interactions between charge and orbitally ordered $d$-electrons are important in many transition metal oxides. We propose an effective energy model for such interactions, parameterized with DFT+U calculations, so that energy contributions of both electronic and lattice origin can be simultaneously accounted for. The model is applied to the low-temperature phase of magnetite, for which we propose a new ground state structure. The effective interactions on the B-lattice of Fe$_3$O$_4$ can be interpreted in terms of electrostatics and short-range Kugel-Khomskii exchange coupling. The frustration between optimal charge and orbital orderings leads to a complex energy landscape whereby the supercell for the charge ordering, orbital ordering and
ionic displacements can all be different.
\end{abstract}

\maketitle

\section{Introduction}
The physical properties of transition-metal oxides (TMO) are determined by two degrees of freedom that are often closely related:
the quantum state of interacting electrons, and the ionic position and/or motion. By itself the description of correlated
$d$-electrons, if bound to or nearly localized on specific transition metal sites, is already complex. The magnetic, charge (electron or hole) and orbital (when the $d$ shell of the transition metal ion is partially filled) degrees of freedom of localized $d$-electrons are coupled via electrostatic, direct and/or super-exchange interactions (for a
review, see e.g.\ refs.\ \onlinecite{Kugel1975SPSS285,Tokura2000S462,Dagotto2005S257}). The interplay between these degrees of freedom and their possible ordering play an important role in understanding such phenomena as metal-insulator transitions, high-temperature superconductivity and colossal magnetoresistance. On the other hand, the ionic displacements may mediate
Jahn-Teller interactions of degenerate orbitals and induce orbital ordering (OO) (see e.g.\ ref.\ \onlinecite{Gehring1975RPP1}). The energetic effects of
both mechanisms are often of the same order of magnitude ($\sim 10-10^2$ meV). It is therefore of
theoretical and practical interest to investigate the combined electronic and lattice effects on the electronic ground state.

First-principles calculations based on the density functional theory (DFT) provide a natural way to incorporate both electronic and ionic degrees of freedom in ``real-world'' materials. Since such methods provide direct information only about the energy of the system as a whole, the microscopic relationship between the involved degrees of freedom has to be extracted indirectly.



Recently, we showed how the effective interactions of localized $d$-electrons (minority spin
$d$-state $t_{2g}^1 e_{g}^0$ of high-spin Fe$^{2+}$) and holes (high-spin Fe$^{3+}$) in the mixed-valence oxide Li$_x$FePO$_4$
could be extracted from first-principles calculations \cite{Zhou2006PRL155704}.
Since two types
of charge carriers, Li-ions and $d$-electrons, coexist in this material, our energy model includes the ionic, ion-electron
interactions as well.
The inter-site coupling parameters could be obtained with the cluster expansion approach \cite{Sanchez1984PA334} whereby an Ising-like
Hamiltonian in electron occupation variables is fitted to DFT+U \cite{Anisimov1991PRB943} total energy calculations of
different charge ordering (CO) and ionic ordering patterns. It was found that the effective electron interactions exhibited
strong electrostatic character at short range and lattice influence at long range. The accuracy of this approach was supported by
the good agreement between the computed \cite{Zhou2006PRL155704} and experimental \cite{Delacourt2005NM254,Dodd2006ESSL151}  temperature-composition phase diagram.
Since the $t_{2g}$ degeneracy in Li$_x$FePO$_4$ is lifted by the irregular FeO$_6$ octahedra, we did not explicitly
consider the orbital degree of freedom for Fe$^{2+}$ in that work.

In a system with degenerate localized $d$ or $f$ states, multiple self-consistent Kohn-Sham solutions may appear, a reflection of the existence of orbital ordering. Since $f$ electrons are well localized, calculations are often trapped in local minima, making it difficult to finding the ground state (for very recent discussions, see ref.\ \cite{Zhou2009PRB125127}). In comparison, the strong crystal field effects in $d$-systems make the calculations relatively straightforward. The computational methods and settings are presented in section \ref{sec:method}, the obtained results and relevant discussions in section \ref{sec:results}, and finally the conclusions and outlook for future work in section \ref{sec:conclusions}.

\section{Computational method} \label{sec:method}
\subsection{Energy model}
In this paper we explore an approach to derive orbital physics from DFT total energies.
A model is introduced for first-principles determination of the effective interactions of {\it charge} and
{\it orbital} ordered (COO) electrons, and applied to \fo. 
The model includes electrostatics, lattice distortion etc in a consistent manner. 
Consider a general energy expression:  
\begin{equation} \label{eq:Potts}
E[\vec{\epsilon}]= E_0 + E_{i}(\epsilon_i) +  E_{ij}(\epsilon_i , \epsilon_j) +  E_{ijk}(\epsilon_i , \epsilon_j, \epsilon_k ) +
\cdots
\end{equation}
where $\epsilon_i$ represents the electronic state (hole and/or orbital) on site $i$ and $\vec\epsilon$ is the system's
configuration of states.
Summation over repeating indices are implied.
The point term $E_{i}$ describes the electron chemical potential and splitting of the orbitals. Besides the pair interaction
matrices $E_{ij}$, one includes in general higher order contributions, e.g.\ $E_{ijk}$.
In practice this model may be too general to use.
Preempting the finding that quantum effects that distinguish the different orbitals vanish at long distance, it is more
convenient to separate orbital-independent interactions $E_c$ from orbital-dependent $E_o$, attributed to the charge and the
orbital degrees of freedom, respectively. The former can be described by a binary (electron/hole) cluster expansion model
\cite{Sanchez1984PA334,Zhou2006PRL155704}, which captures both short and long-range effects \cite{Zhou2006PRL155704}. Therefore we rewrite
eq.~\ref{eq:Potts} as
\begin{eqnarray} \label{eq:CE-Potts}
E[\vec{\epsilon}]&=& E_c[\vec{\epsilon}] + E_o[\vec{\epsilon}]\\
E_c[\vec{\epsilon}]&=& E_c[\vec{\tilde\epsilon}] =
 J_{\emptyset} + J_i \tilde\epsilon_i +
 J_{ij} \tilde\epsilon_i \tilde\epsilon_j +
 \cdots \nonumber \\
E_o[\vec{\epsilon}] &=& V_{i} (\epsilon_i )  + V_{ij} (\epsilon_i , \epsilon_j) + \cdots \nonumber
\end{eqnarray}
where $J$'s are effective cluster interactions (ECI) of a cluster expansion \cite{Sanchez1984PA334}, $\tilde\epsilon = -1$ when $\epsilon$ is the hole
state and +1 otherwise, and $V$ designates the residual orbital-dependent interactions. In a real material, we expect
$V$ to vanish more quickly with distance than orbital-independent $J$.
The charge ordering energy
$E_c[\vec{\epsilon}]= E_c[\vec{\tilde\epsilon}]$ is function of charge configuration $\vec{\tilde\epsilon}$ alone.
Eq.~\ref{eq:CE-Potts} allows for consistent treatment of charge and orbital interactions.
As will be discussed in our example, the separation into charge and orbital contributions is not unique and depends on our choice
of independent parameters. If chosen appropriately, the parameters $J$ and $V$ can provide useful physical insight.
They combine to derive $E[\vec{\epsilon}]$, which is always meaningful.
We will also see that lattice symmetry can further simplify eq.~\ref{eq:CE-Potts}. 

\subsection{Magnetite Structure}
We will concentrate on magnetite \fo, a mixed-valence oxide with nominal iron valence between 2+ and 3+. At room temperature \fo\
has inverted cubic spinel structure $Fd\bar{3}m$ with tetrahedral A sites occupied by one-third of the cations as Fe$^{3+}$, and
octahedral B sites by two-thirds of the cations with nominal valence 2.5+. At $T_V \sim 120$ K it undergoes the Verwey transition
lowering the symmetry \cite{Verwey1939N327,Verwey1947JCP181}. Although the very existence of B-site Fe$^{2+}$/Fe$^{3+}$ ordering at
low-temperature (low-T) is not completely agreed upon, with some experimental data arguing against it \cite{Garcia2001PRB54110,Subias2004PRL156408} and some supporting it \cite{Wright2001PRL266401,Wright2002PRB214422,Goff2005JPCM7633,Nazarenko2006PRL56403,Bimbi2008PRB45115}, we note
that recent theoretical \cite{Leonov2004PRL146404,Jeng2004PRL156403,Leonov2006PRB165117, Jeng2006PRB195115} and experimental
\cite{Huang2006PRL96401,Schlappa2008PRL26406, Lorenzo2008PRL226401} results advocate both charge and orbital ordering in \fo.
The low-T structure has also been studied with model Hamiltonian \cite{Seo2002PRB85107,Uzu2006JPSJ43704}. Piekarz and co-workers discussed the interplay of the electronic and ionic degrees of freedom in explaining the mechanism of the Verwey transition with first-principles and group theoretical arguments \cite{Piekarz2006PRL156402,*Piekarz2007PRB165124}.

\begin{figure}
\includegraphics[width=0.8 \linewidth]{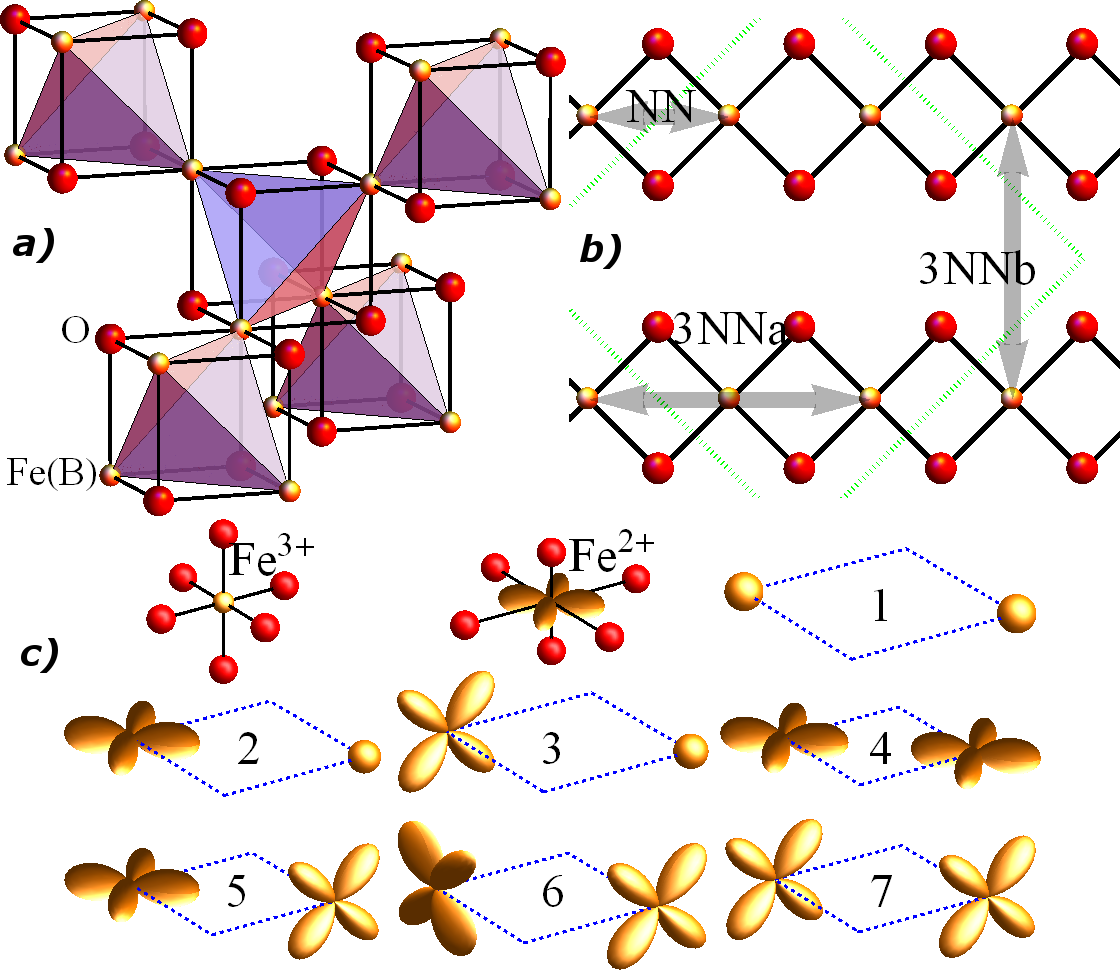}
\caption{(Color online) a) Fragment of the \fo\ structure with Fe(B)-O cubes, and the pyrochlore lattice formed by Fe(B) tetrahedra. b) One
$\{001\}$ plane of Fe atoms in $\langle 110\rangle$
chains. Three distinct B pairs, two within a chain and one between two neighboring chains, are highlighted.
c) Seven distinct $t_{2g}$ orbital interactions between coplanar Fe$^{2+}$ or Fe$^{3+}$ ions. Bond length difference in Fe$^{2+}$ is
exaggerated. Sphere indicates Fe$^{3+}$.
} \label{fig:structure}
\end{figure}
Fig.~\ref{fig:structure}a shows the local environment of the Fe(B) atoms, where bonded Fe(B) and O atoms form corner-sharing cubes.
The corner-sharing FeO$_6$ octahedra align as chains in the $\langle 110\rangle$ directions (we always refer to the fcc cell
coordinates). Two parallel chains are shown in fig.~\ref{fig:structure}b with nearest-neighbor (NN) and third-NN (3NN) B-sites
highlighted. There are two distinct 3NN pairs: 3NNa on one chain with an intermediate Fe atom and 3NNb across two chains with no middle atom. The 2NN Fe(B) atoms do not share a $\{001\}$ plane and are not shown.
The B-sites form a pyrochlore lattice. Anderson found that the frustrated NN interaction on this lattice leads to highly
degenerate ground states with each B-tetrahedron occupied by two Fe$^{2+}$ and two Fe$^{3+}$ \cite{Anderson1956PR1008}. The low-T CO
structure proposed by Verwey \cite{Verwey1939N327,Verwey1947JCP181} (for a picture see e.g.\ fig.~1 of ref.\ \onlinecite{Jeng2006PRB195115}) satisfies
the Anderson condition, while some recent CO models, e.g.\ in refs.~\onlinecite{Zuo1990PRB8451,Wright2001PRL266401,Wright2002PRB214422}, do not. To
our understanding the low-temperature structure is still not fully resolved \cite{Garcia2004JPCM145,Walz2002JPCM285}. The low-T structure and
the charge (and orbital) energetics therefore invites quantitative study. Here we present a detailed study of COO in \fo\ for the
dual purpose of exploring first-principles calculation of orbital interactions in general, and to try to better understand the
structure and origin of low-T phase.

Magnetite is ferrimagnetic ($T_c \approx 860$ K) with antiparallel magnetic moments on the A and B sites at low-T. We fix the
magnetic configuration as such in this work and focus on the charge and orbital degrees of freedom. The FeO$_6$ crystal field splits
the five minority spin $d$-states into three $t_{2g}$ states and two higher-energy $e_{g}$ states. At low temperature four states are
accessible at each B site: the Fe$^{3+}$ hole and the Fe$^{2+}$ with $t_{2g}$ orbitals $xy$, $yz$, and $xz$ (see fig.~\ref{fig:structure}c).
The symmetry and three-fold $t_{2g}$ degeneracy reduces the number of independent $V$'s.
We show in fig.~\ref{fig:structure}c symmetrically distinct elements of the orbital interaction matrix $V$ for NN (or
3NNa,b) pairs. Eq.~\ref{eq:CE-Potts} is simplified as follows: The electron chemical potential term $J_i$ is unnecessary as
the number of electrons is fixed in stoichiometric \fo.
The orbital point energy $V_{i}$ is dropped due to the $t_{2g}$ degeneracy.
Some of the $V$ matrix elements are linearly dependent on the $J$ terms and may be removed. For example, $V(1)$, orbital interaction
type 1 (hole-hole) is already represented by the charge interaction $J\tilde{\epsilon}_i\tilde{\epsilon}_j$ at the same sites.
Only three matrix elements in fig.~\ref{fig:structure}c are linearly independent within eq.~\ref{eq:CE-Potts}. We choose to
keep $V(4)$, $V(5)$ and $V(7)$ and take $V(6)$ as reference. The orbital-independent $J$ is then unambiguously defined as between
the reference orbital states, and $V$ is the adjustment to $J$ when the electronic states are not the reference. For example, the
total interaction between NN $xy$ electrons on the $ab$-plane is $J_{\mathrm NN}+ V_{\mathrm NN}(4)$.

\subsection{Computational details}
To parameterize the simplified eq.~\ref{eq:CE-Potts} we have performed Generalized Gradient Approximation (GGA) + Hubbard $U$
(GGA+U) \cite{Anisimov1991PRB943} calculations at $U_{\mathrm{eff}}\equiv U-J=4$ eV (unless otherwise stated).
All calculations were carried out using the VASP package \cite{Kresse1996PRB11169,Kresse1999PRB1758} with projector augmented wave (PAW) potentials \cite{Blochl1994PRB17953}, energy cutoff of 450 eV, and without any symmetry constraint on ionic and lattice relaxation. Each calculation was initialized in a specific configuration of charge and orbital order and self-consistently converged.
We use supercells of $\frac{1}{\sqrt{2}} \times \frac{1}{\sqrt{2}}
\times 1$,  $\frac{1}{\sqrt{2}} \times \frac{1}{\sqrt{2}} \times 2$, $1 \times 1 \times 1$, $1 \times 1 \times 2$ (designated I,
II, III, IV) relative to the fcc cell (see
fig.~2 of ref.\ \onlinecite{Wright2002PRB214422}, where they were named $P2/m$, $Fd\bar{3}m$, $P2/c$ and $Cc$, respectively)
and $2 \times 1 \times 1$, $4 \times 1 \times 1$ relative to the fcc primitive cell.

The issue or orbital moment is of considerable interest in the electronic structure of Fe$_{3}$O$_{4}$. Despite earlier reports of considerable orbital moment at the B-site Fe$^{2+}$ ions \cite{Huang2004PRL77204}, more recent measurements have found a relatively small orbital/spin moment ratio \cite{Goering2006EL97,Kallmayer2008JAP07D715}. In this work we ignore spin-orbit coupling and assume completely suppressed orbital moment.

The parameters in eq.~\ref{eq:CE-Potts} were
determined with a iterative procedure commonly used in parametrization of cluster expansion: 1) fit GGA+$U$ energy to eq.~\ref{eq:CE-Potts}, 2) search with the obtained parameters for low-energy configurations, and 3) calculate new structures, if any, with GGA+U and go to step 1). This procedure was repeated until the parameters converged and no new ground state emerged. In the end, we calculated 365 distinct COO arrangements. 

\section{Results and discussions} \label{sec:results}

In agreement with refs.~\onlinecite{Leonov2004PRL146404, Jeng2004PRL156403,Leonov2006PRB165117,Jeng2006PRB195115,Pinto2006JPCM10427}, charge
disproportionation of $\lesssim 0.2 e$ is observed between Fe$^{2+}$ and Fe$^{3+}$, with the $t_{2g}$ occupancies in the form of
$\{\alpha, \delta,\delta\}$ or $\{\delta,\delta,\delta\}$ ($\alpha \sim 0.7-1.0$, $\delta \sim 0.0-0.3$), respectively, clearly
validating the notion of separating the Fe ions into distinct valence and orbital states. Another way to distinguish the ions is
via the relaxed Fe-O bond length. Typically the Fe$^{3+}$-O bond is $2.03\pm 0.04$ \AA, while the six Fe$^{2+}$-O bonds are
$2.11$ \AA\ on average, with four elongated bonds of $\sim 2.15$ \AA\ within, and two shorter bonds perpendicular to, the plane
of occupied orbital (see fig.~\ref{fig:structure}c), proving that Fe$^{2+}$ is a Jahn-Teller active ion in \fo.
Our result confirms previous assessment \cite{Leonov2004PRL146404, Jeng2004PRL156403}
that the Fe$^{2+}$ can be understood as Jahn-Teller active small polaron.

\begin{figure}
\includegraphics[width=0.75 \linewidth]{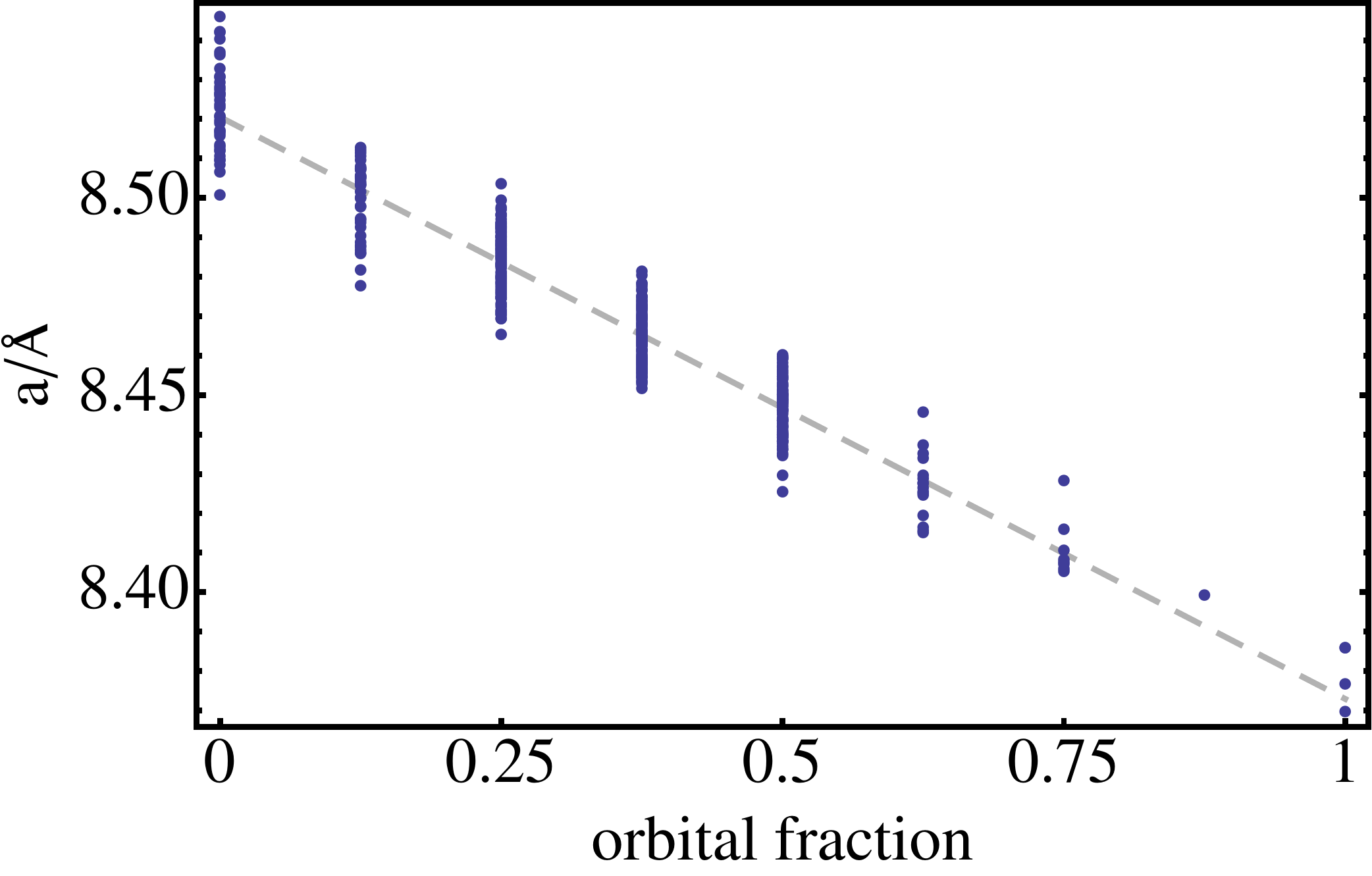}
\caption{(Color online) Lattice parameter versus the fraction of ``perpendicular'' orbitals. The gray dashed line is the best linear fit.
} \label{fig:latticeparameter}
\end{figure}

The effect of orbital order on crystal structure is clearly demonstrated in fig.~\ref{fig:latticeparameter}, where the lattice parameters (e.g.\
$c$) of 128 relaxed structures in the $1 \times 1 \times 1$ fcc cell are shown as function of $f$, the fraction  of ``perpendicular'' (e.g.\ $xy$ as opposed to $c$)
orbitals among all $t_{2g}$ orbitals of B-site Fe$^{2+}$. Therefore a random configuration corresponds to $f=1/3$. A linear fitting (dashed line in fig.~\ref{fig:latticeparameter}) of $(8.521- 0.148 f)$ \AA\ well captures the overall trend of the lattice parameters.
The variation $\sim 0.15$ \AA\ across the range of $f$ is considerably larger than the variation $\sim 0.04$ \AA\ at fixed $f$.

\begin{figure}
\includegraphics[width=0.75 \linewidth]{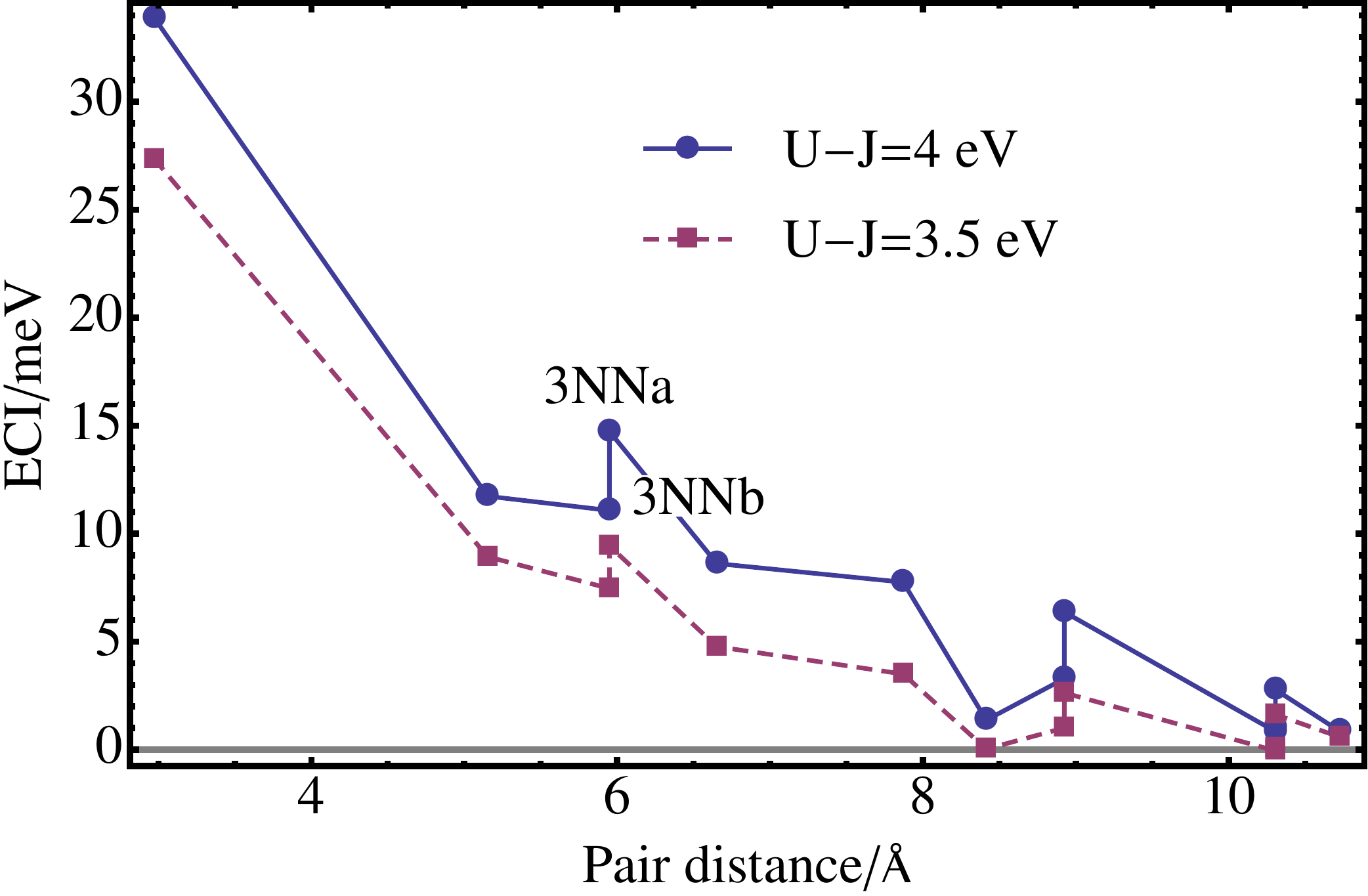}
\caption{(Color online) Orbital-independent pair ECI versus pair distance at $U_{\mathrm{eff}}=4$ eV (solid line) and 3.5 eV (dashed line). 
} \label{fig:paireci}
\end{figure}
The best fit of the GGA+U energies to eq.~\ref{eq:CE-Potts} has a cross validation score \cite{Walle2002JPE348} of 6 meV and root-mean-square (RMS) error of 5 meV per
formula unit (f.u.), with 27 parameters. The orbital-independent $J$'s include the constant term, 3 small triplet and 2 small
quadruplet terms, and most significantly, 13 pair interactions shown in fig.~\ref{fig:paireci}. Note that these are {\em effective}
interactions including the many physical effects: electrostatics, screening, relaxation, covalency, etc. The NN pair ECI (solid
line) is the largest orbital independent interaction (34 meV), reflecting strong electrostatic repulsion. The orbital independent
$J$'s weaken considerably with distance and fall below 1 meV at $ \gtrsim 10$ \AA. A similar trend was observed in Li$_x$FePO$_4$
\cite{Zhou2006PRL155704}. 
\begin{table}[htbp]
\begin{ruledtabular}
\begin{tabular}{c|ccc|ccc}
 & \multicolumn{3}{c|}{$U_{\mathrm{eff}}=4$ eV} & \multicolumn{3}{c}{$U_{\mathrm{eff}}=3.5$ eV}\\
Pair   & $V(4)$ & $V(5)$ & $V(7)$ &  $V(4)$ & $V(5)$ & $V(7)$ \\
\hline
  NN & 106 & 30 & 3 & 125 & 33 & 3 \\
  3NNa & 32 & 10 & -10 & 38 & 8 & -5 \\
  3NNb & 6 & 0 & 0 & 7 & 0 & 0 \\
\end{tabular}
\end{ruledtabular}
\caption{Orbital interaction parameters $V$ (fig.~\ref{fig:structure}c) in meV at three Fe(B) pairs in fig.~\ref{fig:structure}b.
 \label{tab:orbitalinteractions}}
\end{table}

The orbital interactions $V(n)$ ($n=4,5,7$) for NN and 3NNa,b are listed in table~\ref{tab:orbitalinteractions}. $V_{\mathrm
NN}(4)$ is by far the largest, which can be understood from 1) 
the $\sigma$-bond-like orientation resulting in strong antiferromagnetic exchange interaction between same-spin electrons and 2)
unfavorable quadrupole interactions. The orbital interactions obey the Kugel-Khomskii model \cite{Kugel1975SPSS285}, with $V(6)=0$
(reference) being the most stable. The 3NNa interactions are in general weaker than the NN, while the 3NNb and 2NN (not shown)
are even smaller than 3NNa, though the distance is the same or shorter. Considering the
different topology, the weak yet appreciable 3NNa orbital couplings suggest that the $\langle 110 \rangle$ chains of Fe/O atoms may transmit exchange
beyond NN.
 Note that a full interpretation of these effective interactions should
include not only electronic but also lattice effects, e.g.\ Jahn-Teller coupling.

\begin{figure}
\includegraphics[width=0.6 \linewidth]{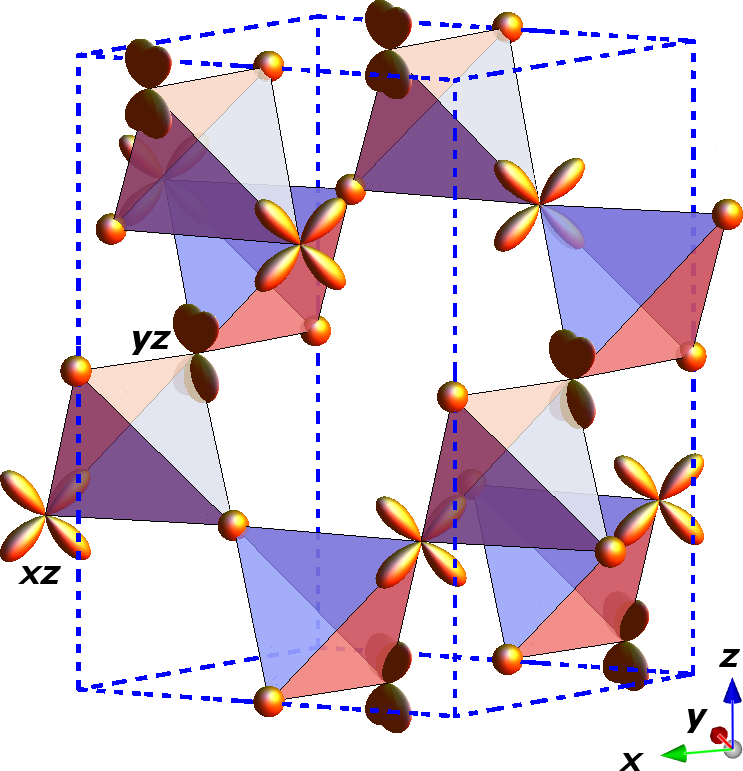}
\caption{(Color online) The ground state COO pattern found in this work within supercell I ($\frac{1}{\sqrt{2}} \times \frac{1}{\sqrt{2}} \times 1$ relative to fcc)
} \label{fig:low-E-config}
\end{figure}

\begin{table}[htbp]
\begin{ruledtabular}
\begin{tabular}{l|llll|l|c}
 & \multicolumn{4}{c|}{GGA+U E in supercells I-IV} & &  Anderson \\
Structure & $\frac{1}{\sqrt{2}}  \frac{1}{\sqrt{2}}  1$ &
$\frac{1}{\sqrt{2}}  \frac{1}{\sqrt{2}}  2$ & $1  1  1$ & $1  1  2$  & $E_o^m$ & rule met? \\
\hline
Verwey
& 0 (ref.) 
&  \multicolumn{1}{l}{-9} 
& -11 
& \multicolumn{1}{l|}{-11}  
& 10 & Y \\
Fig.~\ref{fig:low-E-config}
& \multicolumn{1}{l}{-30} 
& -28 
& -45 
& \multicolumn{1}{l|}{-45} 
& 20 & Y \\
$Cc$ \cite{Jeng2006PRB195115} & & &
& -26 
& 7 & N \\
$P2/c$ \cite{Jeng2006PRB195115} &
& -4 
&
&
& 11 & N   %
\end{tabular}
\end{ruledtabular}
\caption{Calculated energy and predicted optimal orbital energy $E_o^m$ (eq.~\ref{eq:min-E}) in meV/f.u.\ of certain structures.
The first two structures are calculated with OO optimized in different supercells (designated relative to fcc), and the last two
with COO patterns from ref.~\onlinecite{Jeng2006PRB195115}. The last column indicates whether the Anderson condition is met.
 \label{tab:energy-supercell}}
\end{table}

To facilitate discussions we define the optimal orbital energy $E^m_o$ of a given charge pattern $\vec{\tilde\epsilon}$ minimized
over OO's compatible with $\vec{\tilde\epsilon}$.
\begin{equation} \label{eq:min-E}
E_o^m[\vec{\tilde\epsilon}]= \min_{\vec{\epsilon} \in \vec{\tilde\epsilon}} E_o[\vec{\epsilon}].
\end{equation}
The optimal orbital pattern is then defined as the one minimizing eq.~\ref{eq:min-E}. The Anderson degeneracy of the charge part
of the energy landscape $E_c$, present with only NN interactions, is lifted by longer range charge interactions.
The configurational space size for $N$ f.u.\ is $C^{2N}_{N}$. On the other hand, the search for the optimal orbital
energy $E_o^m[\vec{\tilde\epsilon}]$ of a given CO $\vec{\tilde\epsilon}$ is also frustrated in a space of $3^N$.
The orbital energy $E_o^m[\vec{\tilde\epsilon}]$ is also complicated by NN and longer range orbital interactions. Given the
complex energy landscape, we use the above parameters to search the COO configuration space by enumeration in supercells II and
III, and with Monte Carlo-based methods in larger supercells . The ground state {\it COO pattern} (fig.~\ref{fig:low-E-config}) we find has
the periodicity of $\frac{1}{\sqrt{2}} \times \frac{1}{\sqrt{2}} \times 1$ (though the periodicity of the {\it structure} is
larger; see later discussions). As shown in fig.~\ref{fig:low-E-config}, the structure has equal number of Fe$^{2+}$/Fe$^{3+}$ on each $ab$
plane and uniform $xz$ or $yz$ electrons on alternate planes, i.e.\ no charge but orbital modulation along $c$. We list the GGA+U
energy of four structures, including our ground state, in table~\ref{tab:energy-supercell}. For the first two structures, the OO
was optimized in supercells I-IV to study their periodicity. First, consider the original Verwey CO model with alternate $ab$
planes of electrons and holes, i.e.\ charge modulation along $c$. A large enough supercell is needed to find the optimal OO of
the Verwey CO model. In supercell I, all electrons occupy the $xy$ state
, 
while in larger cells II-IV, they equally occupy $xy$ and $xz$ to lower the energy by 9$\sim$11 meV (the variation is due to small
convergence error in different supercells). Our ground state structure (fig.~\ref{fig:low-E-config}) is confirmed to have the lowest GGA+U
energy among all of our calculations. It has the same optimal OO in the four supercells, but lowers its energy by $15$ meV in
supercell III or IV compared to I or II, a difference too large to be a convergence error. It is found that in supercells I and
II the Fe$^{2+}$-O bond lengths have similar distribution with standard deviation of 0.047 \AA,  while in cells III and IV the
deviation is 0.063 \AA. We believe that the energy difference has to do with lattice coupling of Jahn-Teller active Fe$^{2+}$:
even with the same COO configuration the ionic positions in a small supercell are more constrained, reducing the chances of
cooperative distortions.
The structure therefore has the periodicity of supercell III, with space group $P4_1$. 
Lastly, two previously proposed COO candidates \cite{Jeng2006PRB195115}, with space group $Cc$
\cite{Zuo1990PRB8451} and $P2/c$ \cite{Wright2001PRL266401}, 
respectively, are included. Both are less stable than our ground state. We confirm the conclusion of Jeng {\it et al.} that $Cc$
is more stable than $P2c$ \cite{Jeng2006PRB195115}. For the charge pattern of the $Cc$ structure, an OO 2 meV lower than the one
reported in ref.~\onlinecite{Jeng2006PRB195115} is found.
The predicted optimal orbital energy $E_o^m$ (eq.~\ref{eq:min-E}) of our structure is 20 meV with all NN Fe$^{2+}$-Fe$^{2+}$
interactions of the unfavorable type 5. Both $P2/c$ and $Cc$ have relatively small $E_o^m$ of
about 10 meV, since many of their NN interactions are of type 6 or 7. 

The experimental low-temperature structure of magnetite is not yet known clearly enough to compare with, but the fact that our ground state has different
unit cell and space group than found in experiment \cite{Wright2001PRL266401} means we have not completely resolved the problem.
Nevertheless, several observations can be made from our results. First, like the charge energy, the orbital energy is also
frustrated. For example, the optimal OO for each of the four structures in table~\ref{tab:energy-supercell} include unfavorable
orbital interactions $V_{NN}(n)$ ($n=5,7$).
Secondly, the charge and orbital energies are competing: a structure may have low charge energy $E_c$ or orbital energy $E_o^m$,
but not both. Taking the Anderson condition as an approximate indicator of low $E_c$, our ground state COO has low $E_c$ but
relatively large $E_o^m$, while the $Cc$ and $P2/c$ are the opposite (table~\ref{tab:energy-supercell}). Thirdly, the frustrated,
competing interactions make the ground state search sensitive to the interaction parameters. It is possible that our search
failed to find a ground state in a larger supercell because of the numerical sensitivity. Other possibilities for the incomplete agreement with experimental supercell might be the missing physics not described by our calculations. This includes spin fluctuations beyond the assumption of a fixed magnetic configuration, and may also include
fluctuations between complex structures of close energies. Lastly, our results illustrate some
possible mechanisms that can break the cubic symmetry and form the low-T GS structure: (1) charge order as Verwey originally
proposed, (2) charge {\it and} orbital order as exemplified in the Verwey CO model whose COO supercell is larger than the CO
supercell, and (3) lattice coupling of Fe$^{2+}$ ions  as seen in our structure (fig.~\ref{fig:low-E-config}), the periodicity of which,
decided by the arrangement of Jahn-Teller distortions, is larger than that of the COO.

To evaluate the impact of the Hubbard term $U_{\mathrm{eff}}$ in our first-principles approach, we have calculated 300 structures with a
smaller $U_{\mathrm{eff}}=3.5$ eV. The pair
ECIs for $U_{\mathrm{eff}}=3.5$ eV are shown as the dashed line in fig.~\ref{fig:paireci}. They are slightly reduced compared to $U_{\mathrm{eff}}=4$
eV, mainly because the largely electrostatic ECIs
 scale as $(\Delta q)^2$ where $\Delta q$ is the charge difference between the 2+/3+ ions. With
smaller $U_{\mathrm{eff}}$ the $d$-electrons become more delocalized and $\Delta q$ generally decreases
\cite{Zhou2004PRB201101,Wenzel2007PRB214430}. As shown in table~\ref{tab:orbitalinteractions} the
orbital interactions $V$ are relatively stable yet some are notably larger at $U_{\mathrm{eff}}=3.5$ eV. Presumably the reason is that the
exchange integral, sensitive to the spatial distribution of the wavefunctions, increases with
delocalization.  It is also possible that the mathematical separation in
eq.~\ref{eq:CE-Potts} of charge and orbital terms is not physically complete, and there is some compensation in $J$ and
$V$ with varying $U_{\mathrm{eff}}$. The smaller Hubbard parameter does not considerably change the results in
table~\ref{tab:energy-supercell} and related discussions.

\section{Conclusions} \label{sec:conclusions}
In this work we have attempted to describe the charge and orbital degrees of freedom in \fo\ with a classical effective energy
model. Electronic and lattice effects are both included through first-principles calculated energies from which the model is
parametrized. The calculated charge and orbital interactions in \fo\ are found to be physically meaningful. The energy landscape
is complex in terms of frustrated charge and orbital interactions as well as their competition. Additionally, although our
predicted ground state structure has smaller periodicity than experimentally observed, it reveals the possibility that not only
charge and orbital ordering, but the Jahn-Teller lattice distortions may also decide the structure. Therefore this work may help
better understand the problem of the low-T magnetite structure. Beyond magnetite, our approach can be easily adapted to explore
other transition metal oxides where charge and/or orbital order exist.

This work is supported by DOE under contract DE-FG02-96ER45571, and in part by NSF through the National Partnership for Advanced
Computing Infrastructure using SDSC Datastar. FZ thanks Dr.\ C.\ Fischer for his help in data analysis.


%

\end{document}